\documentclass[%
 reprint,
 amsmath,amssymb,
 aps,
pra,
]{revtex4-1}

\usepackage{graphicx}
\usepackage{dcolumn}
\usepackage{bm}

\begin{document}

\title{Single-image geometric-flow x-ray speckle tracking}

\author{David M. Paganin}
\affiliation{School of Physics and Astronomy, Monash University, Victoria 3800, Australia}
\author{H\'{e}l\`{e}ne Labriet}
\affiliation{Novitom, 3 av doyen Louis Weil 38000 Grenoble, France}
\affiliation{Universit\'{e} Grenoble Alpes, Rayonnnement Synchrotron et Recherche M\'{e}dicale, F-38000 Grenoble, France}
\author{Emmanuel Brun}
\affiliation{Universit\'{e} Grenoble Alpes, Rayonnnement Synchrotron et Recherche M\'{e}dicale, F-38000 Grenoble, France}
\author{Sebastien Berujon}
\affiliation{European Synchrotron Radiation Facility, F-38043 Grenoble, France}

\date{\today}

\begin{abstract}
We develop a speckle-tracking method for x-ray phase-contrast imaging, based on the concept of geometric flow. This flow is a conserved current associated with deformation of illuminating x-ray speckles induced by passage through a sample. The method provides a rapid, efficient, and accurate algorithm for quantitative phase imaging. It is highly photon efficient and able to image dynamic objects, since a single radiograph of the sample is sufficient for the phase recovery. We experimentally quantify the resolution and contrast of the approach with both two-dimensional and three-dimensional phase-imaging applications using x-ray synchrotron radiation. Finally, we discuss adaptations of the method to imaging with compact x-ray sources that have a large source size and significant spectral bandwidth.
\end{abstract}

\maketitle

\section{Introduction}

Increasing contrast in x-ray imaging is of fundamental importance in many scientific fields such as material sciences, cultural heritage, and medical imaging. X-ray phase contrast imaging (XPCI) \cite{wilkins1996} significantly increases the contrast between materials and tissues of very close composition, as, for instance, in distinguishing tumorous from healthy biological soft tissue \cite{Wang2014c,Zhao2012}. While conventional x-ray imaging is based on the local attenuation of a photon beam, XPCI is sensitive to the real part of the complex optical refractive index of a material, which is responsible for light refraction. Despite significant advances in the field, XPCI remains costly at large scale facilities and very challenging on conventional x-ray sources since these latter experimental systems require high stability and precision optics \cite{endrizzi2018}.

Near-field speckle-based x-ray phase-imaging techniques appeared recently, and were quickly demonstrated attractive with respect to other XPCI techniques, on account of their simplicity of experimental implementation \cite{berujon2012,morgan2012}. See \citet{zdora2018} for a comprehensive recent review. Unlike other modulation-based methods, a simple object with small random features is used to generate the modulating speckle observable in the bright field of an x-ray beam. In the Fresnel regime of hard x rays, speckles resemble in size the object generating them over large distances, and such “near-field” speckles can be tracked between different images taken at different points in time or locations in space. Furthermore, spatially random intensity modulation can be generated either from interference effects of the light scattered from a speckle mask containing small randomly distributed grains— the resulting intensity structures are in that case true speckle when such speckles are fully developed—or by absorption contrast from a mask with randomly distributed attenuating apertures. Such features have made speckle-based methods readily compatible with low coherence sources, enabling them to rapidly spread beyond synchrotrons, to be demonstrated applicable with laboratory sources \cite{Zanette2014}. In parallel, scientists have developed various speckle-tracking processing methods, usually requiring several acquisitions, to optimize the technique’s sensitivity and resolution while also accessing the so-called dark-field signal \cite{berujon2015}.

In spite of their various advantages and successful applications, the requirement for several sample exposures to achieve high resolution raises strong challenges on the sample and speckle mask positioning reproducibility, which are, for instance, beyond what is acceptable when imaging living patients or dynamic samples.

Herein, we propose to solve this drawback using a speckle-tracking approach that is based on optical energy conservation and geometric flow. With respect to other speckle-tracking phase-contrast techniques, the method described here intrinsically senses both lens and derivative terms of the phase. These simultaneously accounted-for terms are associated with propagation-induced phase contrast and differential phase contrast, respectively. Moreover, the method given here implicitly rather than explicitly tracks speckles. The implementation of the method is fast, robust, and extremely efficient. These attractive aspects are experimentally illustrated using reconstructions from data collected at an x-ray synchrotron. Finally, we discuss future development of the method, whose potential adaptation to laboratory x-ray sources is highly promising.

\section{Theory}

\begin{figure}
\includegraphics[width=0.45\textwidth,scale=0.14]{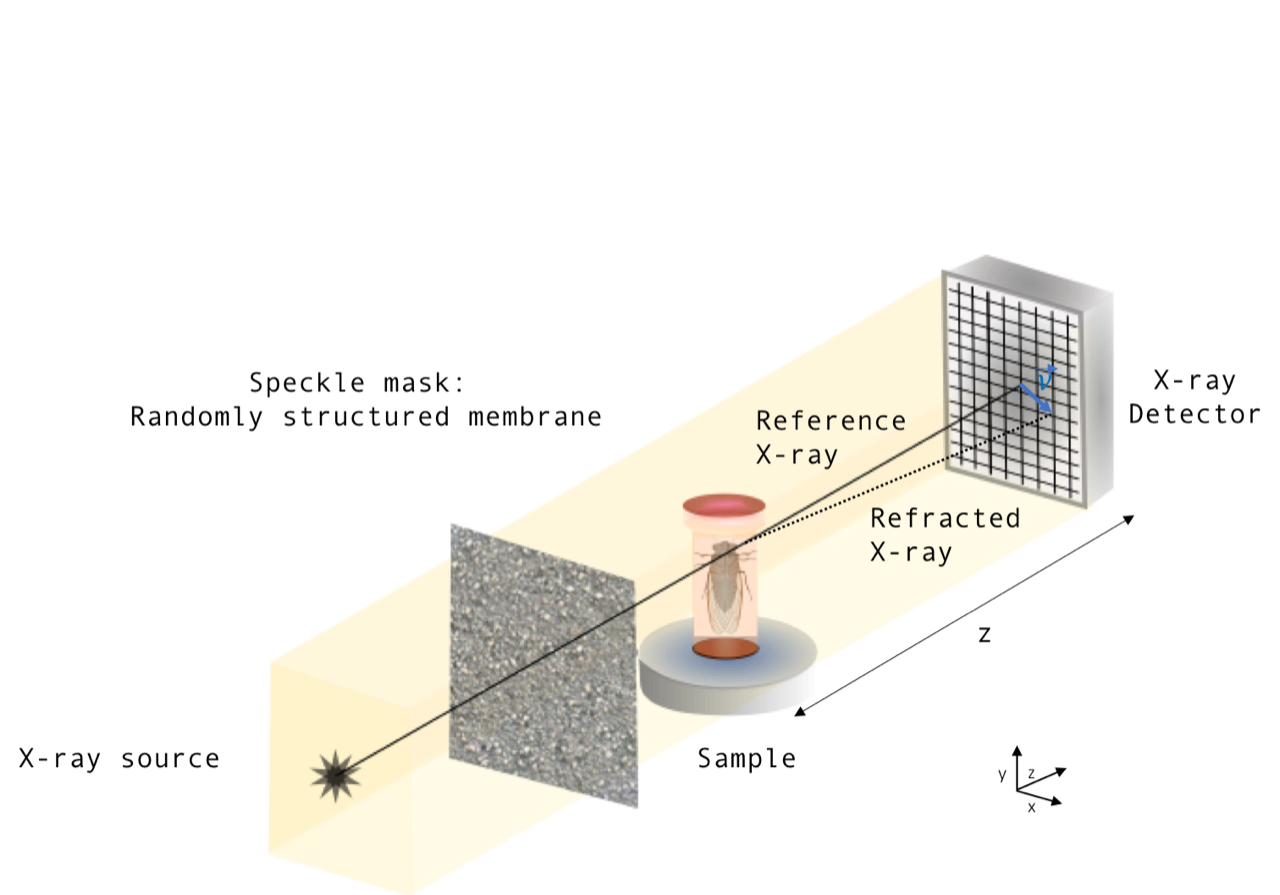}
\caption{Generic setup for x-ray-speckle tracking}
\label{fig:SpeckleSetUp}
\end{figure}

Consider Fig.~\ref{fig:SpeckleSetUp}, where paraxial forward-propagating x rays illuminate a speckle mask, before passing through a thin object and then traversing a distance $Z$ to the planar surface of a position-sensitive detector. Let $I_R(x,y)$ be the reference speckle, i.e., the image taken in the absence of the object, where $(x,y)$ are Cartesian coordinates in the plane perpendicular to the optic axis $z$. The image in the presence of the sample is denoted by $I_S(x,y)$. Assume the sample to be a thin perfectly x-ray-transparent object, whose presence geometrically distorts the reference x-ray speckles. Assume this geometric distortion to conserve the integrated intensity of the reference speckle image, both locally and globally. Hence one can describe the deformation of $I_R(x,y)$ into $I_S(x,y)$ as a geometric flow with a conserved current, namely a transverse flow of intensity that obeys the continuity equation. This flow corresponds to the continuous warping of the intensity distribution in the $z=Z>0$ plane perpendicular to the optic axis, effected by continuously evolving the speckled intensity distribution over the plane $z=Z$ in the absence of the object, into the intensity distribution over the plane $z=Z$ in the presence of the object.

The above considerations permit us to write
\begin{equation}
I_R(x,y)-I_S(x,y)\approx\nabla_{\perp}\cdot[I_R(x,y){\bf D}_{\perp}(x,y)],
\label{eq:1}
\end{equation}
\noindent where $\nabla_{\perp}$ denotes the gradient operator in the $(x,y)$ plane, and ${\bf D}_{\perp}(x,y)=(D_x(x,y),D_y(x,y))$ is the displacement field that distorts each feature in the reference image $I_R(x,y)$ into the corresponding feature in the image $I_S(x,y)$ taken in the presence of the sample. Note that expansion of the right side of Eq.~\ref{eq:1} gives a ``prism term'' $\nabla_{\perp}I_R(x,y)\cdot{\bf D}_{\perp}(x,y)$ corresponding to the transverse motion of speckles in each of two orthogonal directions, in a contribution to the signal which is amplified by the presence of strong intensity gradients such as are provided by an illuminating speckle field.  To this prism term is added a ``lensing term'' $I_R(x,y)\nabla_{\perp}\cdot{\bf D}_{\perp}(x,y)$ corresponding to local concentration or rarefaction of intensity. Modelling the intensity flow via Eq.~\ref{eq:1} takes both effects into account, implying the method developed below to simultaneously utilize both differential phase contrast and propagation-based phase contrast.

Following \citeauthor{teague1983} \cite{teague1983} and \citeauthor{paganin1998} \cite{paganin1998}, assume that the flow—associated with deforming $I_R(x,y)$ into $I_S(x,y)$—has a transverse current density proportional to $I_R(x,y){\bf D}_{\perp}(x,y)$ that may be written as the gradient of a scalar auxiliary function $\Lambda(x,y)$. Hence,
\begin{equation}
I_R(x,y){\bf D}_{\perp}(x,y)\approx \nabla_{\perp}\Lambda(x,y).
\label{eq:2}
\end{equation}
\noindent This approximation amounts to neglecting the curl of a vector potential which would otherwise need to be added to the right side of the above equation, in the Helmholtz decomposition of the vector field on the left. Stated differently, we have assumed the previously mentioned geometric flow to be a gradient flow. Physically, this amounts to the assumption that the angular-momentum density of the flow is much smaller in magnitude than its linear-momentum density ({\em cf.}~\citet{schmalz2011}).

Our auxiliary function transforms Eq.~\ref{eq:1} into:
\begin{equation}
I_R(x,y)-I_S(x,y)=\nabla_{\perp}^2\Lambda (x,y).
\label{eq:3}
\end{equation}
\noindent Since the left side is known from measurement data, one may solve this Poisson equation using a variety of numerical methods (e.g.,~multigrid methods, finite-element methods, relaxation methods, etc.). Appropriate boundary conditions may be either measured or known \emph{a priori}. For example, if the object is entirely immersed within the field of view of the illuminating speckle field, zero Dirichlet boundary conditions (and/or zero Neumann boundary conditions) may be assumed. While we implicitly assume this case below, such an assumption is easily relaxed.

We now give a method for solving Eq.~\ref{eq:1}, which has some parallels with that derived by Paganin and Nugent \cite{paganin1998,paganin2006} in a different context. Fourier transform Eq.~\ref{eq:3} with respect to $x$ and $y$, then use the Fourier derivative theorem, to give
\begin{equation}
\mathcal{F}[I_R(x,y)-I_S(x,y)]=-(k_x^2+k_y^2) \mathcal{F} [\Lambda (x,y)].
\label{eq:4}
\end{equation}
Here, $\mathcal{F}$ denotes Fourier transformation with respect to $x$ and $y$, $(k_x,k_y)$ are the corresponding Fourier coordinates, and we have used the Fourier-transform convention from \citet{paganin2006}.  Solving for $\Lambda (x,y)$ then gives:
\begin{equation}
\Lambda (x,y) = \mathcal{F}^{-1}\left\{ \frac{\mathcal{F}[I_S(x,y)-I_R(x,y)]}{k_x^2+k_y^2} \right\}.
\label{eq:5}
\end{equation}
The division-by-zero Fourier-space singularity, which corresponds to an irrelevant constant offset in the auxiliary function, implies that the point $(k_x,k_y)=(0,0)$ at the origin of Fourier space should be excluded from the above expression. This amounts to taking the Cauchy principal value of the integrals which must be performed in evaluating the above expression. Numerically, one simply omits the zero-frequency pixel in Fourier space, from the domain of integration.

The Fourier derivative theorem implies that the transverse gradient operator may be written as \cite{paganin2006}:
\begin{equation}
\nabla_{\perp}=i\mathcal{F}^{-1}(k_x,k_y)\mathcal{F} ,
\label{eq:6}
\end{equation}
\noindent where all operators act from right to left. Apply $\nabla_{\perp}$ to both sides of Eq.~\ref{eq:5}; then use Eq.~\ref{eq:6} on the right-hand side and Eq.~\ref{eq:2} on the left-hand side; finally, divide both sides of the resulting expression by $I_R(x,y) > 0$, to give
\begin{eqnarray}
{\bf D}_{\perp}(x,y) \quad\quad\quad\quad\quad\quad\quad\quad\quad\quad\quad\quad\quad\quad\quad\quad\quad\quad\quad\quad
\label{eq:7}
\\ \nonumber = \frac{i}{I_R(x,y)} \mathcal{F}^{-1} \left( (k_x,k_y) \left\{ \frac{\mathcal{F}[I_S(x,y)-I_R(x,y)]}{k_x^2+k_y^2} \right\} \right).
\end{eqnarray}

In analogy with the concept of a velocity potential introduced by Lagrange into potential flow theory for classical irrotational fluids, assume the displacement field of the flow---in the mapping deforming $I_R(x,y)$ into $I_S(x,y)$---to be irrotational. This allows us to write the displacement field as the gradient of a scalar potential $d(x,y)$:
\begin{eqnarray}
{\bf D}_{\perp}(x,y)\equiv(D_x(x,y),D_y(x,y)) \approx \nabla_{\perp}d(x,y).
\label{eq:8}
\end{eqnarray}
\noindent We can obtain $d(x,y)$ from ${\bf D}_{\perp}(x,y)$ using the Fourier transform-based algorithm \cite{arnison2004,kottler2007,huang2015}:
\begin{eqnarray}
d(x,y)=\mathcal{F}^{-1}\left\{ \frac{\mathcal{F}[D_x(x,y)+iD_y(x,y)]}{ik_x-k_y}\right\}.
\label{eq:9}
\end{eqnarray}

Thus far we have not utilized any particular assumptions related to optical imaging, implying significant generality in the preceding development. The physical reason underpinning this generality is the fact that the local flows we consider are conserved currents embodying a local conservation principle (via the continuity equation), a setting that is far more generally applicable than a particular differential equation governing a particular nondissipative conserved field. Notwithstanding the desirability of the level of generality with which we have hitherto worked, we now utilize some assumptions pertinent to speckle-tracking using hard x rays. We restrict attention to this case for the remainder of the paper.

For paraxial quasimonochromatic complex scalar x-ray radiation with wavelength $\lambda=2\pi/k$ corresponding to wavenumber k, the geometry of Fig.~\ref{fig:SpeckleSetUp} implies that $d(x,y)$ is related to the transverse phase shift $\phi(x,y)$ which the thin nonabsorbing sample imparts upon the transmitted x-ray beam:
\begin{eqnarray}
d(x,y)=\frac{Z}{k}\phi(x,y).
\label{eq:10}
\end{eqnarray}
\noindent The transverse gradient of the above expression gives:
\begin{eqnarray}
\nabla_{\perp} d(x,y)=\frac{Z}{k}\nabla_{\perp}\phi(x,y)=Z(\alpha_x(x,y),\alpha_y(x,y)), \quad
\label{eq:11}
\end{eqnarray}
\noindent where $(\alpha_x(x,y),\alpha_y(x,y))$ are the deflection angles that the object imparts on the traversing x-ray beam, in the $x$ and $y$ directions, respectively.

Equations~\ref{eq:10} and \ref{eq:11} allow the preceding very general derivation, for reconstructing the scalar potential $d(x,y)$ and the displacement field ${\bf D}_{\perp}(x,y)$, to be converted into an algorithm for reconstructing the phase shift $\phi(x,y)$ of the object, together with the associated deflection angles $(\alpha_x(x,y),\alpha_y(x,y))$. The formula for reconstructing the phase shift $\phi(x,y)$ is
\begin{eqnarray}
\phi(x,y)=\frac{k}{Z}\,\mathcal{F}^{-1}\left\{\frac{\mathcal{F}[(\hat{\bf x} + i\hat{\bf y})\cdot {\bf D}_{\perp}(x,y)]}{ik_x-k_y}\right\},
\label{eq:final_answer}
\end{eqnarray}
\noindent where $\hat{\bf x}$ and $\hat{\bf y}$ are unit vectors in the $x$ and $y$ directions respectively, and ${\bf D}_{\perp}(x,y)$ is given by Eq.~\ref{eq:7}. While the reconstruction of phase shifts is strictly only meaningful for monochromatic radiation, this requirement may be considerably relaxed when one instead reconstructs deflection angles. These angles $(\alpha_x(x,y),\alpha_y(x,y))$ may be obtained  directly from Eq.~\ref{eq:7} via:
\begin{eqnarray}
\label{eq:final_answer_angles}
(\alpha_x(x,y),\alpha_y(x,y))=\frac{{\bf D}_{\perp}(x,y)}{Z}.
\end{eqnarray}

Note that the magnitude of the Fourier-space filter, in the denominator of Eqs.~\ref{eq:9} and \ref{eq:final_answer}, is the inverse of the Ramachandran-Lakshminarayanan (“Ram--Lak”) filter. Hence if our speckle-tracking method is combined with tomography, the filter in braces in Eq.~\ref{eq:9} or \ref{eq:final_answer} will be multiplied by $(k_x^2+k_y^2)^{1/2}$. The fact that $(k_x^2+k_y^2)^{1/2}/(ik_x-k_y)=\exp[i\Phi(k_x,k_y)]$ where $\Phi(k_x,k_y)$ is a polar angle in Fourier space, implies that in speckle-tracking tomography using our method, one can simply omit the Ram-Lak filter and replace division by $(ik_x-k_y)$ with multiplication by the vortical unit-modulus function $\exp[i\Phi(k_x,k_y)]$. This makes the x-ray speckle-tracking tomography much more local, and more stable, than filtered backprojection. Note also that the assumption that ${\bf D}_{\perp}(x,y)$ be irrotational [see Eq.~\ref{eq:8}] is exact if $\phi(x,y)$ is continuous and single valued, a condition that is guaranteed if the projection approximation \cite{paganin2006} is valid and the sample’s refractive index is continuous.

\section{Materials and methods}

\begin{figure*}
\includegraphics[width=0.85\textwidth]{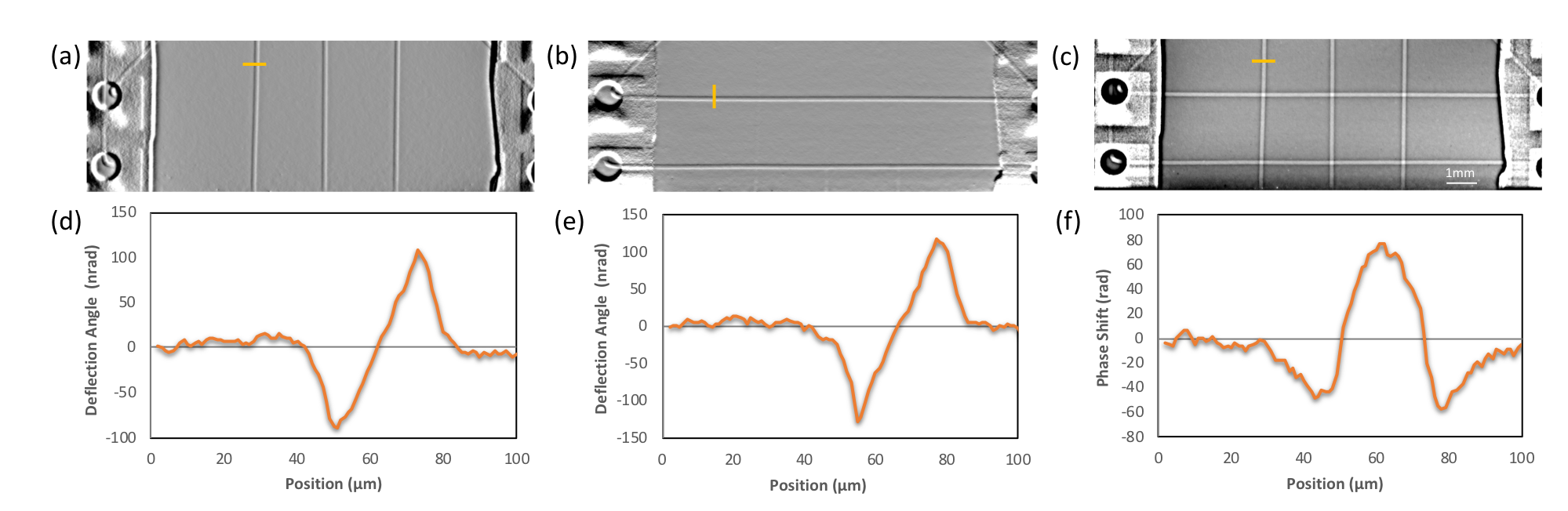}
\caption{Geometric-flow x-ray speckle tracking applied for 2D radiography of 150 $\mu$m diameter nylon wires. Refraction angles measured in the (a) $x$-plane $[\alpha_x(x,y)])$ and (b) $y$-plane $[\alpha_y(x,y)]$. (c) Integrated phase $-\phi(x,y)$. (d)-(f) Profile plots along the segments marked in yellow in (a)-(c) respectively.}
\label{fig:NylonWires}
\end{figure*}

\subsection{Synchrotron x-ray experiments}

Two experiments were conducted at the ID17 Biomedical beamline of the European Synchrotron (ESRF), demonstrating applications of the method.

In the first experiment, a double Si(111) crystal system in Laue--Laue configuration was used to select a quasimonochromatic beam with an energy spread $\frac{ \Delta E}{E} \simeq 10^{-4}$ at the energy $E=52$ keV. Collimation was defined by the natural divergence of the 21-pole wiggler synchrotron source which is about $<$ 1 mrad horizontally and $<$ 0.1 mrad vertically. The geometrical configuration of the experiment resembles Fig.~\ref{fig:SpeckleSetUp} in concept. The x-ray photons first passed through the speckle-generator membrane before traversing the sample located 900 mm away and finally impinging onto the detector placed $Z = 12$ m further downstream. An indirect detection system was used, consisting of a scientific CMOS coupled to magnifying optics to image a scintillator screen made of a gadolinium oxysulfide sheet of 60 μm thickness. The resulting pixel size was $\simeq$ 6 $\mu$m. The images obtained during this experiment, of a homemade phantom, are shown in Fig.~\ref{fig:NylonWires}. The phantom was composed of several orthogonally crossed nylon wires with a 150~$\mu$m diameter.

To study the potential of this speckle-based methodology with a polychromatic beam, a second experiment was carried out using a white synchrotron beam filtered with 0.5 mm of Al and 0.35 mm of Cu. The resulting beam spectrum corresponds to a pink beam with a peak centered at E = 37.3 keV and a spectral bandwidth of approximately 20 keV ($\frac{ \Delta E}{E} \simeq 0.5$). For this experiment, the detector system was the same scientific CMOS camera as before but coupled to different magnifying optics which provided a resulting pixel size of $\simeq$ 3 $\mu$m. The sample was a domestic cicada dried under natural conditions for 10 months after its natural death. The setup configuration was equivalent to the one employed to obtain the images presented in Fig.~\ref{fig:NylonWires}, but the sample-to-detector and the membrane-to-sample distances were set to 4 m and 1 m, respectively. The tomography data consisted of 3000 projections collected periodically during a 360 degree scan of the sample. The center of rotation was transversely off-centred by 200 pixels to operate a so-called half acquisition tomography scan. Such a common procedure allows one to extend the three-dimensional (3D) field of view. The phase images were calculated for each projection using Eqs.~\ref{eq:7} and \ref{eq:final_answer}. The 3D computed tomography (CT) reconstruction was performed using the back-projection implementation described in \citet{mirone2014} and the results are presented in Fig.~\ref{fig:CicadaTomography}.

\subsection{Numerical implementation}

Our algorithm was implemented in PYTHON 3 and is available under the GNU General Public Licence \url{https://github.com/labrieth/spytlab/}~\cite{onlinecode}. The code is not optimized in terms of computation time but provides a readable and understandable implementation of the method. Raw experimental data can be downloaded \url{https://github.com/labrieth/spytlab/}~\cite{onlinecode}.

\section{Results}

\subsection{Radiography\label{Sec:radiography}}

The geometric-flow method for x-ray-speckle tracking was first applied in a two-dimensional (2D) radiography mode to the nylon-wire phantom. The total data consist of one reference speckle image $I_R(x,y)$ in the absence of the sample, and one speckle image IS (x, y) in the presence of the phantom. Figure~\ref{fig:NylonWires} shows 2D maps of the recovered refraction angles [Fig.~\ref{fig:NylonWires}(a)] $\alpha_x(x,y)$, and [Fig.~\ref{fig:NylonWires}(b)] $\alpha_y(x,y)$, together with Fig.~\ref{fig:NylonWires}(c) the recovered phase shifts $\phi(x,y)$.

The figure highlights the method’s quantitativeness for the recovery of phase shifts. Such an aspect allows one to extract the various indexes of refraction composing a sample when using a monochromatic beam, and enable a better distinction of the different materials. The profile plots of the figure present a high signal-to-noise ratio and the standard deviation of the reconstruction error in a region with no sample is below 100 nrad. Such values underline the good stability of the method with respect to noise. Besides, no blurring effect that occurs with most speckle tracking techniques is observed here~\cite{Zdora2017}. In short, the combined high sensitivity and high resolution of the method permits one to image with high fidelity the sample-induced phase shift from a single sample exposure.

\begin{figure*}
\includegraphics[width=0.85\textwidth]{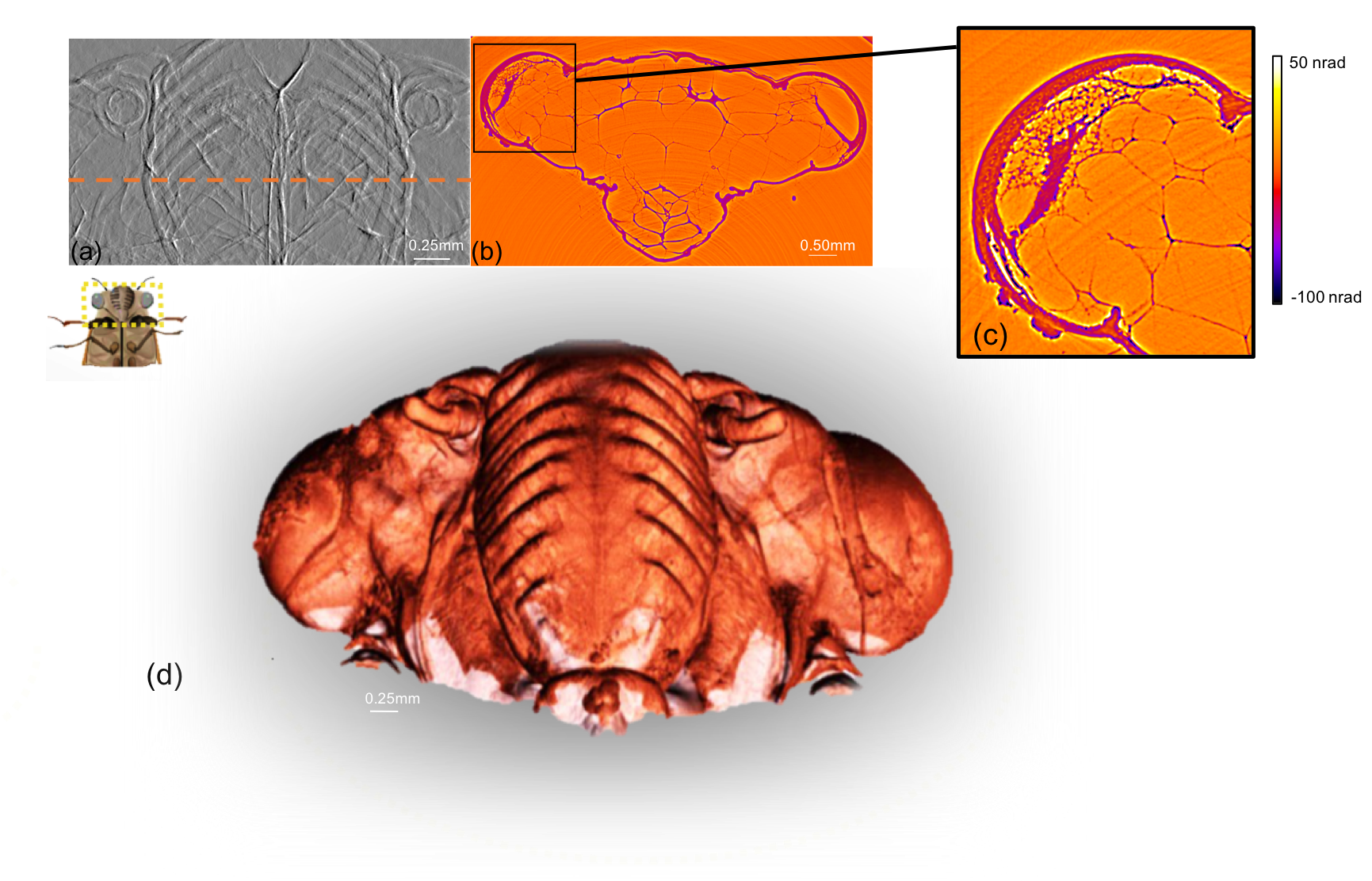}
\caption{CT reconstruction of a cicada head, using geometric-flow x-ray speckle tracking with a single view per projection. (a) Differential phase gradient image obtained for the first projection (horizontal direction) and (b) reconstructed slice image for the cutting plane marked by the horizontal line in the cicada head sketch. (c) Zoom of the compound-eye area marked by the inset in (b) and (d) 3D volume rendering of the cicada head.}
\label{fig:CicadaTomography}
\end{figure*}

\subsection{Computed tomography\label{sec:tomo}}

In a second experiment, we used geometric-flow x-ray speckle tracking to perform phase-contrast tomography of a cicada using a polychromatic x-ray beam. Figure~\ref{fig:CicadaTomography}(a) shows a differential phase gradient projection, Fig.~\ref{fig:CicadaTomography}(b) a CT slice with Fig.~\ref{fig:CicadaTomography}(c) an inset in the compound-eye region and Fig.~\ref{fig:CicadaTomography}(d) a 3D volume rendering. In an attempt to probe the limitations of the method, the data set consisted of only one pair of speckle images [$I_R(x,y)$ and $I_S(x,y)$] for each projection. In practice, $I_R(x,y)$ needs to be measured only once at the beginning or end of the scan and a single image $I_S(x,y)$ is taken for each projection. Despite the reduced total exposure of the sample to the x rays, the method successfully renders the phase gradient images of the cicada head with a limited amount of noise. This noise eventually cancels out in the tomographic reconstruction as one can observe in the reconstructed phase-shift slice (\emph{cf.} our earlier comment regarding the reduction of the Ram--Lak filter and the associated stability for tomography). A zoom-in to the cicada eye region in Fig.~\ref{fig:CicadaTomography}(c) illustrates the high resolution of the image and the subtle details obtained. Thus, the 3D volume reconstruction offers a full visualization of the cicada anatomy at high resolution (antennas, compound eyes, clipeus, and the labrum) which was made possible by the method despite the low absorption of the sample whose content in water is low.

\section{Discussion and Conclusion}

XPCI provides excellent sensitivity for imaging many types of material including biological soft tissues. Nevertheless, numerous challenges remain for phase-contrast imaging to further mature into an effective noninvasive tool with laboratory setups. Indeed, although the methods currently available at laboratories usually exhibit good sensitivity, such performance is balanced by the need for complex experimental setups and/or high stability and high precision optics. Recent developments in the field permit the lifting of certain technological barriers \cite{wang2016,Gromann2017} but many key issues still need to be addressed. For instance, the current methods often remain incompatible with CT in term of dose due to the absorption implied by the use of optical elements.

X-ray near-field speckle tracking is a novel, recently developed XPCI technique sensitive to the first derivative of the phase \cite{berujon2012,morgan2012}. The technique setup besides its simplicity of implementation has the main advantages of having no field-of-view limitation (large diffusers are easy to manufacture), no resolution limitation, and finally the requirements on the beam coherence are low. It is also radiation dose efficient because no absorbing element is used between the sample and the detector meaning that all photons passing through the sample eventually contribute to the image formation. To summarize, the experimental complexity of PCI is translated in x-ray-speckle tracking to only the numerical processing side. Up to now the primary limitations of speckle tracking were that the technique necessitates a large number of sample exposures and time-consuming computer algorithms to achieve a high resolution and sensitivity. The method implemented in this work overcomes these limitations, as we have shown quantitative results using only a single sample exposure.

The method only requires two fast Fourier transforms (FFTs) per projection in order to reconstruct the displacement field ${\bf D}_{\perp}(x,y)$ via Eq.~\ref{eq:7}, or four FFTs per projection if both ${\bf D}_{\perp}(x,y)$ and the phase $\phi(x,y)$ are required [using Eqs.~\ref{eq:7} and \ref{eq:final_answer}]. Again, only one image per projection is needed, once the reference image $I_R(x,y)$ in the absence of the sample has been taken. The method is therefore fast with respect to previously available speckle tracking methods and many other phase sensitive approaches. It is comparable in terms of calculating resources and speed to the widely used propagation-based phase-contrast method of \citet{paganin2002b}, which is now optimized for GPU calculations. Although the long object-to-detector propagation distances of a few meters are difficult to access at other facilities, our results show that in a single image the method is already competitive with other scanning techniques that require multiple images \cite{modregger2011,diemoz2013}. The photon energy used for this experiment also demonstrates the potential of the approach for a range of imaging applications where penetrating x rays or reduced deposited dose are strongly required. The computationally simple reconstruction method of the present paper may be used as a first iterate which is subsequently refined, e.g., if the distortions due to the object are strong, and/or if there is some Fresnel fringing. A variety of methods for such iterative refinement exist, such as those based on conjugate gradients, genetic algorithms, simulated annealing, machine learning, and neural networks.

Given its successful application to data obtained using a broad-band polychromatic beam with energy spread $\frac{ \Delta E}{E} \simeq 0.5$, it is evident that the coherence requirements for the method are rather weak. The method nowhere relies on interferometric phase contrast, but rather only on the transverse redistribution of optical energy on account of the geometric distortion of the speckle field that is induced by its passage through an object. While wavefield phase is referred to in several of the above equations, this always ultimately appears in the form $\nabla_{\perp}\phi/k$, which we know to be a ray deflection angle, from Eq.~\ref{eq:11}. This emphasizes that it is geometric optics and geometric distortion of optical rays that underpin the present formalism, which nowhere relies on interference and which therefore--—as already mentioned--—has limited coherence requirements. It would be interesting in future work to investigate the minimum coherence requirements for the method, with a view to seeing its application to a wide variety of low-brilliance sources.

Speed, simplicity, and breadth of applicability are attractive features of our speckle-tracking method based on the concept
of geometric flow. Since the reference image $I_R(x,y)$  need only be measured once, the method can be easily applied to
dynamic data, where the image $I_S(x,y,t)$ in the presence of the sample is a function of time $t$ . Both the Lagrangian and Eulerian viewpoints for the flow field can be reconstructed. Another direction for future work would be to compare the relative efficacy of a random absorption mask, instead of true interference speckle, being used to generate the single reference
image $I_R(x,y)$ required by the method. We conjecture the method to be relatively insensitive to the nature of the speckle that constitutes $I_R(x,y)$, if this intensity distribution is highly structured spatially. Indeed, often one will have a mixture of both classes of speckle, especially at compact x-ray sources. Speckle methods based on an absorbing mask were demonstrated efficient \cite{wang2016}. Absorption masks are in greater need at high energies where the transverse coherence lengths are smaller and/or with sources of larger size. The idea of self referencing objects, in which the flowing speckle is provided by the sample itself, would also be an interesting avenue for future research, e.g., by tracking the speckles formed when sufficiently coherent x rays pass through lung tissue. Finally, another avenue for future work could be to apply this approach to imaging at different length scales, e.g., for clinical imaging and industrial inspection.

In conclusion, our study shows that the use of a geometric flow approach can address many limitations of existing speckle-based differential x-ray phase-contrast techniques. This technique may be applied to lower-coherence sources such as laboratory setups, thereby spreading its impact beyond synchrotron sources.

\begin{acknowledgments}
In memory of our colleague Claudio Ferrero, whose role was pivotal in bringing the authors of this collaboration together.  We acknowledge useful discussions with Josh Bowden and Claudio Ferrero. Financial support from the Experiment Division of the ESRF for D.M.P. to visit in early 2018 is gratefully acknowledged. E.B. also acknowledges support from LabEx PRIMES (ANR-11-LABX-0063/ANR- 11-IDEX-0007). We thank the ESRF for providing beamtime as well as Alberto Bravin, Alberto Mittone and Herwig Requardt for their technical support. We are grateful to Ludovic Broche for his help during the experiment.
\end{acknowledgments}


\begin{thebibliography}{24}%
\makeatletter
\providecommand \@ifxundefined [1]{%
 \@ifx{#1\undefined}
}%
\providecommand \@ifnum [1]{%
 \ifnum #1\expandafter \@firstoftwo
 \else \expandafter \@secondoftwo
 \fi
}%
\providecommand \@ifx [1]{%
 \ifx #1\expandafter \@firstoftwo
 \else \expandafter \@secondoftwo
 \fi
}%
\providecommand \natexlab [1]{#1}%
\providecommand \enquote  [1]{``#1''}%
\providecommand \bibnamefont  [1]{#1}%
\providecommand \bibfnamefont [1]{#1}%
\providecommand \citenamefont [1]{#1}%
\providecommand \href@noop [0]{\@secondoftwo}%
\providecommand \href [0]{\begingroup \@sanitize@url \@href}%
\providecommand \@href[1]{\@@startlink{#1}\@@href}%
\providecommand \@@href[1]{\endgroup#1\@@endlink}%
\providecommand \@sanitize@url [0]{\catcode `\\12\catcode `\$12\catcode
  `\&12\catcode `\#12\catcode `\^12\catcode `\_12\catcode `\%12\relax}%
\providecommand \@@startlink[1]{}%
\providecommand \@@endlink[0]{}%
\providecommand \url  [0]{\begingroup\@sanitize@url \@url }%
\providecommand \@url [1]{\endgroup\@href {#1}{\urlprefix }}%
\providecommand \urlprefix  [0]{URL }%
\providecommand \Eprint [0]{\href }%
\providecommand \doibase [0]{http://dx.doi.org/}%
\providecommand \selectlanguage [0]{\@gobble}%
\providecommand \bibinfo  [0]{\@secondoftwo}%
\providecommand \bibfield  [0]{\@secondoftwo}%
\providecommand \translation [1]{[#1]}%
\providecommand \BibitemOpen [0]{}%
\providecommand \bibitemStop [0]{}%
\providecommand \bibitemNoStop [0]{.\EOS\space}%
\providecommand \EOS [0]{\spacefactor3000\relax}%
\providecommand \BibitemShut  [1]{\csname bibitem#1\endcsname}%
\let\auto@bib@innerbib\@empty
\bibitem [{\citenamefont {Wilkins}\ \emph {et~al.}(1996)\citenamefont
  {Wilkins}, \citenamefont {Gureyev}, \citenamefont {Gao}, \citenamefont
  {Pogany},\ and\ \citenamefont {Stevenson}}]{wilkins1996}%
  \BibitemOpen
  \bibfield  {author} {\bibinfo {author} {\bibfnamefont {S.~W.}\ \bibnamefont
  {Wilkins}}, \bibinfo {author} {\bibfnamefont {T.~E.}\ \bibnamefont
  {Gureyev}}, \bibinfo {author} {\bibfnamefont {D.}~\bibnamefont {Gao}},
  \bibinfo {author} {\bibfnamefont {A.}~\bibnamefont {Pogany}}, \ and\ \bibinfo
  {author} {\bibfnamefont {A.~W.}\ \bibnamefont {Stevenson}},\ }\href@noop {}
  {\bibfield  {journal} {\bibinfo  {journal} {Nature}\ }\textbf {\bibinfo
  {volume} {384}},\ \bibinfo {pages} {335} (\bibinfo {year}
  {1996})}\BibitemShut {NoStop}%
\bibitem [{\citenamefont {Wang}\ \emph {et~al.}(2014)\citenamefont {Wang},
  \citenamefont {Hauser}, \citenamefont {Singer}, \citenamefont {Trippel},
  \citenamefont {Kubik-Huch}, \citenamefont {Schneider},\ and\ \citenamefont
  {Stampanoni}}]{Wang2014c}%
  \BibitemOpen
  \bibfield  {author} {\bibinfo {author} {\bibfnamefont {Z.}~\bibnamefont
  {Wang}}, \bibinfo {author} {\bibfnamefont {N.}~\bibnamefont {Hauser}},
  \bibinfo {author} {\bibfnamefont {G.}~\bibnamefont {Singer}}, \bibinfo
  {author} {\bibfnamefont {M.}~\bibnamefont {Trippel}}, \bibinfo {author}
  {\bibfnamefont {R.~A.}\ \bibnamefont {Kubik-Huch}}, \bibinfo {author}
  {\bibfnamefont {C.~W.}\ \bibnamefont {Schneider}}, \ and\ \bibinfo {author}
  {\bibfnamefont {M.}~\bibnamefont {Stampanoni}},\ }\href {\doibase
  10.1038/ncomms4797} {\bibfield  {journal} {\bibinfo  {journal} {Nat.
  Commun.}\ }\textbf {\bibinfo {volume} {5}},\ \bibinfo {pages} {3797}
  (\bibinfo {year} {2014})}\BibitemShut {NoStop}%
\bibitem [{\citenamefont {Zhao}\ \emph {et~al.}(2012)\citenamefont {Zhao},
  \citenamefont {Brun}, \citenamefont {Coan}, \citenamefont {Huang},
  \citenamefont {Sztr{\'{o}}kay}, \citenamefont {Diemoz}, \citenamefont
  {Liebhardt}, \citenamefont {Mittone}, \citenamefont {Gasilov}, \citenamefont
  {Miao},\ and\ \citenamefont {Bravin}}]{Zhao2012}%
  \BibitemOpen
  \bibfield  {author} {\bibinfo {author} {\bibfnamefont {Y.}~\bibnamefont
  {Zhao}}, \bibinfo {author} {\bibfnamefont {E.}~\bibnamefont {Brun}}, \bibinfo
  {author} {\bibfnamefont {P.}~\bibnamefont {Coan}}, \bibinfo {author}
  {\bibfnamefont {Z.}~\bibnamefont {Huang}}, \bibinfo {author} {\bibfnamefont
  {A.}~\bibnamefont {Sztr{\'{o}}kay}}, \bibinfo {author} {\bibfnamefont
  {P.~C.}\ \bibnamefont {Diemoz}}, \bibinfo {author} {\bibfnamefont
  {S.}~\bibnamefont {Liebhardt}}, \bibinfo {author} {\bibfnamefont
  {A.}~\bibnamefont {Mittone}}, \bibinfo {author} {\bibfnamefont
  {S.}~\bibnamefont {Gasilov}}, \bibinfo {author} {\bibfnamefont
  {J.}~\bibnamefont {Miao}}, \ and\ \bibinfo {author} {\bibfnamefont
  {A.}~\bibnamefont {Bravin}},\ }\href {\doibase 10.1073/pnas.1204460109}
  {\bibfield  {journal} {\bibinfo  {journal} {Proc. Nat. Acad. Sci.}\ }\textbf
  {\bibinfo {volume} {109}},\ \bibinfo {pages} {18290} (\bibinfo {year}
  {2012})}\BibitemShut {NoStop}%
\bibitem [{\citenamefont {Endrizzi}(2018)}]{endrizzi2018}%
  \BibitemOpen
  \bibfield  {author} {\bibinfo {author} {\bibfnamefont {M.}~\bibnamefont
  {Endrizzi}},\ }\href@noop {} {\bibfield  {journal} {\bibinfo  {journal}
  {Nucl. Instr. Meth. Phys. Res. A}\ }\textbf {\bibinfo {volume} {878}},\
  \bibinfo {pages} {88} (\bibinfo {year} {2018})}\BibitemShut {NoStop}%
\bibitem [{\citenamefont {Berujon}\ \emph {et~al.}(2012)\citenamefont
  {Berujon}, \citenamefont {Ziegler}, \citenamefont {Cerbino},\ and\
  \citenamefont {Peverini}}]{berujon2012}%
  \BibitemOpen
  \bibfield  {author} {\bibinfo {author} {\bibfnamefont {S.}~\bibnamefont
  {Berujon}}, \bibinfo {author} {\bibfnamefont {E.}~\bibnamefont {Ziegler}},
  \bibinfo {author} {\bibfnamefont {R.}~\bibnamefont {Cerbino}}, \ and\
  \bibinfo {author} {\bibfnamefont {L.}~\bibnamefont {Peverini}},\ }\href@noop
  {} {\bibfield  {journal} {\bibinfo  {journal} {Phys. Rev. Lett.}\ }\textbf
  {\bibinfo {volume} {108}},\ \bibinfo {pages} {158102} (\bibinfo {year}
  {2012})}\BibitemShut {NoStop}%
\bibitem [{\citenamefont {Morgan}\ \emph {et~al.}(2012)\citenamefont {Morgan},
  \citenamefont {Paganin},\ and\ \citenamefont {Siu}}]{morgan2012}%
  \BibitemOpen
  \bibfield  {author} {\bibinfo {author} {\bibfnamefont {K.~S.}\ \bibnamefont
  {Morgan}}, \bibinfo {author} {\bibfnamefont {D.~M.}\ \bibnamefont {Paganin}},
  \ and\ \bibinfo {author} {\bibfnamefont {K.~K.~W.}\ \bibnamefont {Siu}},\
  }\href@noop {} {\bibfield  {journal} {\bibinfo  {journal} {Appl. Phys.
  Lett.}\ }\textbf {\bibinfo {volume} {100}},\ \bibinfo {pages} {124102}
  (\bibinfo {year} {2012})}\BibitemShut {NoStop}%
\bibitem [{\citenamefont {Zdora}(2018)}]{zdora2018}%
  \BibitemOpen
  \bibfield  {author} {\bibinfo {author} {\bibfnamefont {M.-C.}\ \bibnamefont
  {Zdora}},\ }\href@noop {} {\bibfield  {journal} {\bibinfo  {journal} {J.
  Imaging}\ }\textbf {\bibinfo {volume} {4}},\ \bibinfo {pages} {60} (\bibinfo
  {year} {2018})}\BibitemShut {NoStop}%
\bibitem [{\citenamefont {Zanette}\ \emph {et~al.}(2014)\citenamefont
  {Zanette}, \citenamefont {Zhou}, \citenamefont {Burvall}, \citenamefont
  {Lundstr{\"{o}}m}, \citenamefont {Larsson}, \citenamefont {Zdora},
  \citenamefont {Thibault}, \citenamefont {Pfeiffer},\ and\ \citenamefont
  {Hertz}}]{Zanette2014}%
  \BibitemOpen
  \bibfield  {author} {\bibinfo {author} {\bibfnamefont {I.}~\bibnamefont
  {Zanette}}, \bibinfo {author} {\bibfnamefont {T.}~\bibnamefont {Zhou}},
  \bibinfo {author} {\bibfnamefont {A.}~\bibnamefont {Burvall}}, \bibinfo
  {author} {\bibfnamefont {U.}~\bibnamefont {Lundstr{\"{o}}m}}, \bibinfo
  {author} {\bibfnamefont {D.}~\bibnamefont {Larsson}}, \bibinfo {author}
  {\bibfnamefont {M.}~\bibnamefont {Zdora}}, \bibinfo {author} {\bibfnamefont
  {P.}~\bibnamefont {Thibault}}, \bibinfo {author} {\bibfnamefont
  {F.}~\bibnamefont {Pfeiffer}}, \ and\ \bibinfo {author} {\bibfnamefont
  {H.}~\bibnamefont {Hertz}},\ }\href {\doibase 10.1103/PhysRevLett.112.253903}
  {\bibfield  {journal} {\bibinfo  {journal} {Phys. Rev. Lett.}\ }\textbf
  {\bibinfo {volume} {112}},\ \bibinfo {pages} {253903} (\bibinfo {year}
  {2014})}\BibitemShut {NoStop}%
\bibitem [{\citenamefont {Berujon}\ and\ \citenamefont
  {Ziegler}(2015)}]{berujon2015}%
  \BibitemOpen
  \bibfield  {author} {\bibinfo {author} {\bibfnamefont {S.}~\bibnamefont
  {Berujon}}\ and\ \bibinfo {author} {\bibfnamefont {E.}~\bibnamefont
  {Ziegler}},\ }\href {\doibase 10.1103/PhysRevA.92.013837} {\bibfield
  {journal} {\bibinfo  {journal} {Phys. Rev. A}\ }\textbf {\bibinfo {volume}
  {92}},\ \bibinfo {pages} {013837} (\bibinfo {year} {2015})}\BibitemShut
  {NoStop}%
\bibitem [{\citenamefont {Teague}(1983)}]{teague1983}%
  \BibitemOpen
  \bibfield  {author} {\bibinfo {author} {\bibfnamefont {M.~R.}\ \bibnamefont
  {Teague}},\ }\href
  {http://www.opticsinfobase.org/abstract.cfm?URI=josa-73-11-1434} {\bibfield
  {journal} {\bibinfo  {journal} {J. Opt. Soc. Am.}\ }\textbf {\bibinfo
  {volume} {73}},\ \bibinfo {pages} {1434} (\bibinfo {year}
  {1983})}\BibitemShut {NoStop}%
\bibitem [{\citenamefont {Paganin}\ and\ \citenamefont
  {Nugent}(1998)}]{paganin1998}%
  \BibitemOpen
  \bibfield  {author} {\bibinfo {author} {\bibfnamefont {D.}~\bibnamefont
  {Paganin}}\ and\ \bibinfo {author} {\bibfnamefont {K.~A.}\ \bibnamefont
  {Nugent}},\ }\href {https://link.aps.org/doi/10.1103/PhysRevLett.80.2586}
  {\bibfield  {journal} {\bibinfo  {journal} {Phys. Rev. Lett.}\ }\textbf
  {\bibinfo {volume} {80}},\ \bibinfo {pages} {2586} (\bibinfo {year}
  {1998})}\BibitemShut {NoStop}%
\bibitem [{\citenamefont {Schmalz}\ \emph {et~al.}(2011)\citenamefont
  {Schmalz}, \citenamefont {Gureyev}, \citenamefont {Paganin},\ and\
  \citenamefont {Pavlov}}]{schmalz2011}%
  \BibitemOpen
  \bibfield  {author} {\bibinfo {author} {\bibfnamefont {J.~A.}\ \bibnamefont
  {Schmalz}}, \bibinfo {author} {\bibfnamefont {T.~E.}\ \bibnamefont
  {Gureyev}}, \bibinfo {author} {\bibfnamefont {D.~M.}\ \bibnamefont
  {Paganin}}, \ and\ \bibinfo {author} {\bibfnamefont {K.~M.}\ \bibnamefont
  {Pavlov}},\ }\href@noop {} {\bibfield  {journal} {\bibinfo  {journal} {Phys.
  Rev. A}\ }\textbf {\bibinfo {volume} {84}},\ \bibinfo {pages} {023808}
  (\bibinfo {year} {2011})}\BibitemShut {NoStop}%
\bibitem [{\citenamefont {Paganin}(2006)}]{paganin2006}%
  \BibitemOpen
  \bibfield  {author} {\bibinfo {author} {\bibfnamefont {D.~M.}\ \bibnamefont
  {Paganin}},\ }\href@noop {} {\emph {\bibinfo {title} {Coherent X-Ray
  Optics}}},\ Oxford Series on Synchrotron Radiation\ (\bibinfo {address}
  {Oxford},\ \bibinfo {year} {2006})\BibitemShut {NoStop}%
\bibitem [{\citenamefont {Arnison}\ \emph {et~al.}(2004)\citenamefont
  {Arnison}, \citenamefont {Larkin}, \citenamefont {Sheppard}, \citenamefont
  {Smith},\ and\ \citenamefont {Cogswell}}]{arnison2004}%
  \BibitemOpen
  \bibfield  {author} {\bibinfo {author} {\bibfnamefont {M.~R.}\ \bibnamefont
  {Arnison}}, \bibinfo {author} {\bibfnamefont {K.~G.}\ \bibnamefont {Larkin}},
  \bibinfo {author} {\bibfnamefont {C.~J.~R.}\ \bibnamefont {Sheppard}},
  \bibinfo {author} {\bibfnamefont {N.~I.}\ \bibnamefont {Smith}}, \ and\
  \bibinfo {author} {\bibfnamefont {C.~J.}\ \bibnamefont {Cogswell}},\
  }\href@noop {} {\bibfield  {journal} {\bibinfo  {journal} {J. Microsc.}\
  }\textbf {\bibinfo {volume} {214}},\ \bibinfo {pages} {7} (\bibinfo {year}
  {2004})}\BibitemShut {NoStop}%
\bibitem [{\citenamefont {Kottler}\ \emph {et~al.}(2007)\citenamefont
  {Kottler}, \citenamefont {David}, \citenamefont {Pfeiffer},\ and\
  \citenamefont {Bunk}}]{kottler2007}%
  \BibitemOpen
  \bibfield  {author} {\bibinfo {author} {\bibfnamefont {C.}~\bibnamefont
  {Kottler}}, \bibinfo {author} {\bibfnamefont {C.}~\bibnamefont {David}},
  \bibinfo {author} {\bibfnamefont {F.}~\bibnamefont {Pfeiffer}}, \ and\
  \bibinfo {author} {\bibfnamefont {O.}~\bibnamefont {Bunk}},\ }\href
  {http://www.opticsexpress.org/abstract.cfm?URI=oe-15-3-1175} {\bibfield
  {journal} {\bibinfo  {journal} {Opt. Express}\ }\textbf {\bibinfo {volume}
  {15}},\ \bibinfo {pages} {1175} (\bibinfo {year} {2007})}\BibitemShut
  {NoStop}%
\bibitem [{\citenamefont {Huang}\ \emph {et~al.}(2015)\citenamefont {Huang},
  \citenamefont {Idir}, \citenamefont {Zuo}, \citenamefont {Kaznatcheev},
  \citenamefont {Zhou},\ and\ \citenamefont {Asundi}}]{huang2015}%
  \BibitemOpen
  \bibfield  {author} {\bibinfo {author} {\bibfnamefont {L.}~\bibnamefont
  {Huang}}, \bibinfo {author} {\bibfnamefont {M.}~\bibnamefont {Idir}},
  \bibinfo {author} {\bibfnamefont {C.}~\bibnamefont {Zuo}}, \bibinfo {author}
  {\bibfnamefont {K.}~\bibnamefont {Kaznatcheev}}, \bibinfo {author}
  {\bibfnamefont {L.}~\bibnamefont {Zhou}}, \ and\ \bibinfo {author}
  {\bibfnamefont {A.}~\bibnamefont {Asundi}},\ }\href {\doibase
  10.1016/j.optlaseng.2014.07.002} {\bibfield  {journal} {\bibinfo  {journal}
  {Opt. Lasers Eng.}\ }\textbf {\bibinfo {volume} {64}},\ \bibinfo {pages} {1}
  (\bibinfo {year} {2015})}\BibitemShut {NoStop}%
\bibitem [{\citenamefont {Mirone}\ \emph {et~al.}(2014)\citenamefont {Mirone},
  \citenamefont {Brun}, \citenamefont {Gouillart}, \citenamefont {Tafforeau},\
  and\ \citenamefont {Kieffer}}]{mirone2014}%
  \BibitemOpen
  \bibfield  {author} {\bibinfo {author} {\bibfnamefont {A.}~\bibnamefont
  {Mirone}}, \bibinfo {author} {\bibfnamefont {E.}~\bibnamefont {Brun}},
  \bibinfo {author} {\bibfnamefont {E.}~\bibnamefont {Gouillart}}, \bibinfo
  {author} {\bibfnamefont {P.}~\bibnamefont {Tafforeau}}, \ and\ \bibinfo
  {author} {\bibfnamefont {J.}~\bibnamefont {Kieffer}},\ }\href@noop {}
  {\bibfield  {journal} {\bibinfo  {journal} {Nucl. Instrum. Meth. Phys. Res.
  B}\ }\textbf {\bibinfo {volume} {324}},\ \bibinfo {pages} {41} (\bibinfo
  {year} {2014})}\BibitemShut {NoStop}%
\bibitem [{\citenamefont {Paganin}\ \emph {et~al.}(2018)\citenamefont
  {Paganin}, \citenamefont {Labriet}, \citenamefont {Brun},\ and\ \citenamefont
  {Berujon}}]{onlinecode}%
  \BibitemOpen
  \bibfield  {author} {\bibinfo {author} {\bibfnamefont {D.~M.}\ \bibnamefont
  {Paganin}}, \bibinfo {author} {\bibfnamefont {H.}~\bibnamefont {Labriet}},
  \bibinfo {author} {\bibfnamefont {E.}~\bibnamefont {Brun}}, \ and\ \bibinfo
  {author} {\bibfnamefont {S.}~\bibnamefont {Berujon}},\ }\href
  {https://github.com/labrieth/spytlab/} {\enquote {\bibinfo {title}
  {https://github.com/labrieth/spytlab/},}\ } (\bibinfo {year}
  {2018})\BibitemShut {NoStop}%
\bibitem [{\citenamefont {Zdora}\ \emph {et~al.}(2017)\citenamefont {Zdora},
  \citenamefont {Thibault}, \citenamefont {Zhou}, \citenamefont {Koch},
  \citenamefont {Romell}, \citenamefont {Sala}, \citenamefont {Last},
  \citenamefont {Rau},\ and\ \citenamefont {Zanette}}]{Zdora2017}%
  \BibitemOpen
  \bibfield  {author} {\bibinfo {author} {\bibfnamefont {M.-C.}\ \bibnamefont
  {Zdora}}, \bibinfo {author} {\bibfnamefont {P.}~\bibnamefont {Thibault}},
  \bibinfo {author} {\bibfnamefont {T.}~\bibnamefont {Zhou}}, \bibinfo {author}
  {\bibfnamefont {F.~J.}\ \bibnamefont {Koch}}, \bibinfo {author}
  {\bibfnamefont {J.}~\bibnamefont {Romell}}, \bibinfo {author} {\bibfnamefont
  {S.}~\bibnamefont {Sala}}, \bibinfo {author} {\bibfnamefont {A.}~\bibnamefont
  {Last}}, \bibinfo {author} {\bibfnamefont {C.}~\bibnamefont {Rau}}, \ and\
  \bibinfo {author} {\bibfnamefont {I.}~\bibnamefont {Zanette}},\ }\href
  {\doibase 10.1103/PhysRevLett.118.203903} {\bibfield  {journal} {\bibinfo
  {journal} {Phys. Rev. Lett.}\ }\textbf {\bibinfo {volume} {118}},\ \bibinfo
  {pages} {203903} (\bibinfo {year} {2017})}\BibitemShut {NoStop}%
\bibitem [{\citenamefont {Wang}\ \emph {et~al.}(2016)\citenamefont {Wang},
  \citenamefont {Kashyap}, \citenamefont {Cai},\ and\ \citenamefont
  {Sawhney}}]{wang2016}%
  \BibitemOpen
  \bibfield  {author} {\bibinfo {author} {\bibfnamefont {H.}~\bibnamefont
  {Wang}}, \bibinfo {author} {\bibfnamefont {Y.}~\bibnamefont {Kashyap}},
  \bibinfo {author} {\bibfnamefont {B.}~\bibnamefont {Cai}}, \ and\ \bibinfo
  {author} {\bibfnamefont {K.}~\bibnamefont {Sawhney}},\ }\href@noop {}
  {\bibfield  {journal} {\bibinfo  {journal} {Sci. Rep.}\ }\textbf {\bibinfo
  {volume} {6}},\ \bibinfo {pages} {30581} (\bibinfo {year}
  {2016})}\BibitemShut {NoStop}%
\bibitem [{\citenamefont {Gromann}\ \emph {et~al.}(2017)\citenamefont
  {Gromann}, \citenamefont {{De Marco}}, \citenamefont {Willer}, \citenamefont
  {No{\"{e}}l}, \citenamefont {Scherer}, \citenamefont {Renger}, \citenamefont
  {Gleich}, \citenamefont {Achterhold}, \citenamefont {Fingerle}, \citenamefont
  {Muenzel}, \citenamefont {Auweter}, \citenamefont {Hellbach}, \citenamefont
  {Reiser}, \citenamefont {Baehr}, \citenamefont {Dmochewitz}, \citenamefont
  {Schroeter}, \citenamefont {Koch}, \citenamefont {Meyer}, \citenamefont
  {Kunka}, \citenamefont {Mohr}, \citenamefont {Yaroshenko}, \citenamefont
  {Maack}, \citenamefont {Pralow}, \citenamefont {{van der Heijden}},
  \citenamefont {Proksa}, \citenamefont {Koehler}, \citenamefont {Wieberneit},
  \citenamefont {Rindt}, \citenamefont {Rummeny}, \citenamefont {Pfeiffer},\
  and\ \citenamefont {Herzen}}]{Gromann2017}%
  \BibitemOpen
  \bibfield  {author} {\bibinfo {author} {\bibfnamefont {L.~B.}\ \bibnamefont
  {Gromann}}, \bibinfo {author} {\bibfnamefont {F.}~\bibnamefont {{De Marco}}},
  \bibinfo {author} {\bibfnamefont {K.}~\bibnamefont {Willer}}, \bibinfo
  {author} {\bibfnamefont {P.~B.}\ \bibnamefont {No{\"{e}}l}}, \bibinfo
  {author} {\bibfnamefont {K.}~\bibnamefont {Scherer}}, \bibinfo {author}
  {\bibfnamefont {B.}~\bibnamefont {Renger}}, \bibinfo {author} {\bibfnamefont
  {B.}~\bibnamefont {Gleich}}, \bibinfo {author} {\bibfnamefont
  {K.}~\bibnamefont {Achterhold}}, \bibinfo {author} {\bibfnamefont {A.~A.}\
  \bibnamefont {Fingerle}}, \bibinfo {author} {\bibfnamefont {D.}~\bibnamefont
  {Muenzel}}, \bibinfo {author} {\bibfnamefont {S.}~\bibnamefont {Auweter}},
  \bibinfo {author} {\bibfnamefont {K.}~\bibnamefont {Hellbach}}, \bibinfo
  {author} {\bibfnamefont {M.}~\bibnamefont {Reiser}}, \bibinfo {author}
  {\bibfnamefont {A.}~\bibnamefont {Baehr}}, \bibinfo {author} {\bibfnamefont
  {M.}~\bibnamefont {Dmochewitz}}, \bibinfo {author} {\bibfnamefont {T.~J.}\
  \bibnamefont {Schroeter}}, \bibinfo {author} {\bibfnamefont {F.~J.}\
  \bibnamefont {Koch}}, \bibinfo {author} {\bibfnamefont {P.}~\bibnamefont
  {Meyer}}, \bibinfo {author} {\bibfnamefont {D.}~\bibnamefont {Kunka}},
  \bibinfo {author} {\bibfnamefont {J.}~\bibnamefont {Mohr}}, \bibinfo {author}
  {\bibfnamefont {A.}~\bibnamefont {Yaroshenko}}, \bibinfo {author}
  {\bibfnamefont {H.-I.}\ \bibnamefont {Maack}}, \bibinfo {author}
  {\bibfnamefont {T.}~\bibnamefont {Pralow}}, \bibinfo {author} {\bibfnamefont
  {H.}~\bibnamefont {{van der Heijden}}}, \bibinfo {author} {\bibfnamefont
  {R.}~\bibnamefont {Proksa}}, \bibinfo {author} {\bibfnamefont
  {T.}~\bibnamefont {Koehler}}, \bibinfo {author} {\bibfnamefont
  {N.}~\bibnamefont {Wieberneit}}, \bibinfo {author} {\bibfnamefont
  {K.}~\bibnamefont {Rindt}}, \bibinfo {author} {\bibfnamefont {E.~J.}\
  \bibnamefont {Rummeny}}, \bibinfo {author} {\bibfnamefont {F.}~\bibnamefont
  {Pfeiffer}}, \ and\ \bibinfo {author} {\bibfnamefont {J.}~\bibnamefont
  {Herzen}},\ }\href {\doibase 10.1038/s41598-017-05101-w} {\bibfield
  {journal} {\bibinfo  {journal} {Sci. Rep.}\ }\textbf {\bibinfo {volume}
  {7}},\ \bibinfo {pages} {4807} (\bibinfo {year} {2017})}\BibitemShut
  {NoStop}%
\bibitem [{\citenamefont {Paganin}\ \emph {et~al.}(2002)\citenamefont
  {Paganin}, \citenamefont {Mayo}, \citenamefont {Gureyev}, \citenamefont
  {Miller},\ and\ \citenamefont {Wilkins}}]{paganin2002b}%
  \BibitemOpen
  \bibfield  {author} {\bibinfo {author} {\bibfnamefont {D.}~\bibnamefont
  {Paganin}}, \bibinfo {author} {\bibfnamefont {S.~C.}\ \bibnamefont {Mayo}},
  \bibinfo {author} {\bibfnamefont {T.~E.}\ \bibnamefont {Gureyev}}, \bibinfo
  {author} {\bibfnamefont {P.~R.}\ \bibnamefont {Miller}}, \ and\ \bibinfo
  {author} {\bibfnamefont {S.~W.}\ \bibnamefont {Wilkins}},\ }\href {\doibase
  10.1046/j.1365-2818.2002.01010.x} {\bibfield  {journal} {\bibinfo  {journal}
  {J. Microsc.}\ }\textbf {\bibinfo {volume} {206}},\ \bibinfo {pages} {33}
  (\bibinfo {year} {2002})}\BibitemShut {NoStop}%
\bibitem [{\citenamefont {Modregger}\ \emph {et~al.}(2011)\citenamefont
  {Modregger}, \citenamefont {Pinzer}, \citenamefont {Thüring}, \citenamefont
  {Rutishauser}, \citenamefont {David},\ and\ \citenamefont
  {Stampanoni}}]{modregger2011}%
  \BibitemOpen
  \bibfield  {author} {\bibinfo {author} {\bibfnamefont {P.}~\bibnamefont
  {Modregger}}, \bibinfo {author} {\bibfnamefont {B.~R.}\ \bibnamefont
  {Pinzer}}, \bibinfo {author} {\bibfnamefont {T.}~\bibnamefont {Thüring}},
  \bibinfo {author} {\bibfnamefont {S.}~\bibnamefont {Rutishauser}}, \bibinfo
  {author} {\bibfnamefont {C.}~\bibnamefont {David}}, \ and\ \bibinfo {author}
  {\bibfnamefont {M.}~\bibnamefont {Stampanoni}},\ }\href@noop {} {\bibfield
  {journal} {\bibinfo  {journal} {Opt. Express}\ }\textbf {\bibinfo {volume}
  {19}},\ \bibinfo {pages} {18324} (\bibinfo {year} {2011})}\BibitemShut
  {NoStop}%
\bibitem [{\citenamefont {Diemoz}\ \emph {et~al.}(2013)\citenamefont {Diemoz},
  \citenamefont {Endrizzi}, \citenamefont {Zapata}, \citenamefont {Pešić},
  \citenamefont {Rau}, \citenamefont {Bravin}, \citenamefont {Robinson},\ and\
  \citenamefont {Olivo}}]{diemoz2013}%
  \BibitemOpen
  \bibfield  {author} {\bibinfo {author} {\bibfnamefont {P.~C.}\ \bibnamefont
  {Diemoz}}, \bibinfo {author} {\bibfnamefont {M.}~\bibnamefont {Endrizzi}},
  \bibinfo {author} {\bibfnamefont {C.~E.}\ \bibnamefont {Zapata}}, \bibinfo
  {author} {\bibfnamefont {Z.~D.}\ \bibnamefont {Pešić}}, \bibinfo {author}
  {\bibfnamefont {C.}~\bibnamefont {Rau}}, \bibinfo {author} {\bibfnamefont
  {A.}~\bibnamefont {Bravin}}, \bibinfo {author} {\bibfnamefont {I.~K.}\
  \bibnamefont {Robinson}}, \ and\ \bibinfo {author} {\bibfnamefont
  {A.}~\bibnamefont {Olivo}},\ }\href@noop {} {\bibfield  {journal} {\bibinfo
  {journal} {Phys. Rev. Lett.}\ }\textbf {\bibinfo {volume} {110}},\ \bibinfo
  {pages} {138105} (\bibinfo {year} {2013})}\BibitemShut {NoStop}%
\end{thebibliography}

%

%
%

\end{document}